\documentclass[12pt]{iopart}
\usepackage{epsfig,amssymb}
\begin{document}
\title[
Global anisotropy of cosmic rays with space-based detectors
]{
Global anisotropy of arrival directions of ultra-high-energy cosmic
rays: capabilities of space-based detectors
}

\author{
O.\,E.\,Kalashev$^1$,
B.\,A.\,Khrenov$^2$,
P.\,Klimov$^2$,
S.\,Sharakin$^2$
and S.\,V.\,Troitsky$^1$
}

\address{
$^1$Institute for Nuclear Research of the Russian Academy of
Sciences,\\
60th October Anniversary Prospect 7a, 117312, Moscow, Russia\\
$^2$
D.V.~Skobeltsin Institute of Nuclear Physics,\\
M.V.~Lomonosov Moscow State University,
Moscow 119992, Russia
}
\ead{st@ms2.inr.ac.ru}
\begin{abstract}
Planned space-based ultra-high-energy cosmic-ray detectors (TUS,
JEM-EUSO and S-EUSO) are best suited for searches of global
anisotropies in the distribution of arrival directions of cosmic-ray
particles because they will be able to observe the full sky with a single
instrument. We calculate quantitatively the strength of anisotropies
associated with two models of the origin of the highest-energy particles:
the extragalactic model (sources follow the distribution of galaxies in
the Universe) and the superheavy dark-matter model (sources follow the
distribution of dark matter in the Galactic halo). Based on the
expected exposure of the experiments, we estimate the optimal strategy for
efficient search of these effects.
\end{abstract}

\pacno{98.70.Sa}

\maketitle

\section{Introduction}
\label{sec:intro}
The next step in the studies
ultra-high-energy (UHE) cosmic rays (CRs)
is related
to the use of space-based detectors observing the fluorescent light
induced by air showers in the terrestrial atmosphere with large exposures.
It is expected that these instruments will help us to shed light on the
origin of the highest-energy cosmic particles which remains unknown up to
now.

One of the important signatures of particular UHECR models
is the global anisotropy of
arrival directions of the highest-energy events. For instance,
models where the origin of UHECRs is attributed to the
acceleration in astrophysical objects (so-called
``bottom-up'' scenarios) naturally predict that the distribution of arrival
directions follows the distribution of these cosmic accelerators. In the
most common scenario with extragalactic protons and/or nuclei, the
patterns of the distribution of galaxies in the nearby Universe should be
seen~\cite{EGmogila} on the UHECR skymap because of the limited
propagation length of these particles due to the GZK effect~\cite{GZK} or
nuclear photodisintegration. Strong suppression of the cosmic-ray flux is
predicted at highest energies in these models. On the other hand, the
``top-down'' models~\cite{TD} (see Refs.~\cite{TDrev} for reviews and
references) with the distribution of sources in the Galactic halo
following that of the dark matter (which is the case for the superheavy
dark-matter (SHDM) particles and some of topological defects)
predict~\cite{DubTin} the Galactic center-anticenter asymmetry due to the
non-central position of the Sun in the Galaxy (see
Refs.~\cite{DMmogila,ABK1,ABK2} for extensive discussions).
Models of this kind predict continuation of the cosmic-ray spectrum and
gamma-ray dominance beyond $10^{20}$~eV.

Currently, neither the spectrum nor anisotropy observations can
definitely favour one of the two scenarios at the highest energies, $E \sim
10^{20}$~eV. Indeed, the AGASA experiment claims~\cite{AGASAspectrum} the
super-GZK continuation of the spectrum while the HiRes collaboration
reports~\cite{HiRes-cutoff} the observation of the
Greisen--Zatsepin--Kuzmin (GZK) cutoff (data of other experiments,
including recent results of the Pierre Auger Observatory
(PAO)~\cite{PAO-SDspectrum,Engel-CIC}, are not yet conclusive: though the
unsuppressed continuation of the spectrum  is excluded by the Auger data,
the cutoff in the spectrum is not clearly seen). The limits on the
gamma-ray fraction in the primary cosmic-ray flux (currently the most
restrictive ones arise from the AGASA and Yakutsk muon data at
$E>10^{20}$~eV~\cite{gamma}, from the Yakutsk muon data at $E>4 \cdot
10^{19}$~eV~\cite{Yak-gamma} and from the Pierre Auger Observatory data on
the shower geometry at $E>2 \cdot 10^{19}$~eV and
$E>10^{19}$~eV~\cite{PAO-gamma}) disfavour the SHDM scenario and even
exclude it for particular values of the dark-matter parameters (see e.g.\
Ref.~\cite{Risse} for a more detailed discussion of some of these limits).
Current experiments do not report any significant deviations from the
global isotropy at the highest energies\footnote{ After this paper was
submitted, PAO reported a significant deviation from
isotropy~\cite{PAOagn}. The definite interpretation of this effect awaits
further data (see e.g.\ Ref.~\cite{Comment} for discussion). }. This is
however not conclusive both due to the low statistics and due to a limited
field of view of any terrestrial-based installation.

The steeply falling spectrum of cosmic rays makes it very difficult to
obtain a reliable measure of the global anisotropy in {\em any}
combination of terrestrial experiments. Indeed, the relative systematic
difference in the energy estimation between two installations located in
different parts of the Earth (and thus observing different parts of the
sky) can hardly be made smaller than some $\sim 15\%$. Such a relative
error would give $\sim 30\%$ higher integral flux seen by one of the
experiments with respect to the other at the same reconstructed energy.
Thus, possible observations of global anisotropy can be attributed both to
a physical effect and to unknown systematics in the energy determination.
Moreover, a seemingly isotropic distribution of the arrival directions
over the sky might represent a physically anisotropic one masked by the
systematic effects.

On the other hand, the planned space-based experiments, e.g.\
TUS~\cite{TUS}, JEM-EUSO~\cite{EUSO} or S-EUSO~\cite{S-EUSO}, will provide
a
unique opportunity to observe full sky with a single
detector. While not being free from systematic uncertainties in the energy
determination, an experiment of this kind would not introduce
strong direction-dependent systematics and thus would be able to perform
the studies of the global anisotropy at high confidence.

One may expect that in future space-based experiments with their huge
exposures, particular sources of UHECR will be determined on
event-by-event basis for the case of the astrophysical scenario. This is
not an easy task, however: limited
angular resolution together with large numbers of events
would lead
to identification problems similar to thoose of the gamma-ray
astronomy\footnote{See e.g.\
Refs.~\cite{EGRET-statistics} for
discussions of statistical methods of analysis of photon-by-photon EGRET
data. Currently, the gamma-ray sky at energies $(\sim 0.1 \div 1)$~GeV
contains 101 identified source, 170 unidentified ones and a strong
non-uniform diffuse background of unidentified origin.} but strongly
enhanced due to magnetic deflections of the charged cosmic-ray particles.
The searches for global patterns in the distribution of UHECR arrival
directions will thus be important in any case.

A robust method for the study of any global asymmetry in the arrival
directions is the harmonic analysis (see e.g.\ Ref.~\cite{Sommers}). It
works perfectly if the predicted effect may be clearly seen in the first
few harmonics but requires large statistics to reveal/exclude more
complicated patterns. Here, we determine the optimal strategy for the
searches of global anisotropies even with the low-resolution experiments
(TUS) and for short exposure times.

The strategy is to fix two regions of the sky (not necessary covering full
$4\pi$) which are expected to provide the strongest contrast  in
over/underdensity of events with respect to the null hypothesis of the
isotropic distribution. The shape and the size of these regions, as well
as the energy range of the events, are determined {\it a priory} in order to
balance the strength of the expected anisotropy (increasing for smaller
regions and higher energies) and its statistical significance. The aim of
this paper is to simulate optimal regions and energies for TUS, JEM-EUSO
and S-EUSO, suitable to test the scenarios of extragalactic sources and of
decays of superheavy dark matter concentrated in the Galactic halo. To this
end, we perform new and improved (with respect to previous works)
simulations of the distribution of arrival directions expected in both
cases.

The paper is organised as follows. In Sec.~\ref{sec:EG} and
Sec.~\ref{sec:DM}, we discuss our simulations for extragalactic sources
and for SHDM decays, respectively. Sec.~\ref{sec:experiments} gives
specific predictions for three coming spaceborn experiments, TUS, JEM-EUSO
and S-EUSO. Sec.~\ref{sec:concl} contains our brief conclusions. Some
technical details are collected in \ref{app:density-function}.

\section{The ``bottom-up'' scenario: sources follow the distribution
of extragalactic luminuous matter}
\label{sec:EG}
A number of astrophysical sources were suggested where acceleration of
cosmic rays up to the highest energies can take place (see
Refs.~\cite{sources} for reviews and summary). A common assumption is that
the distribution of the sources follows that of luminuous matter, that is
of galaxies. We should stress that this approach does not imply that all
galaxies emit cosmic rays of UHE energies -- this is certainly ruled out;
instead, it assumes that the number density of the sources is,
on average, proportional to the number density of the galaxies.
This latter assumption is true for most of the suggested cosmic
accelerators (active galactic nuclei, starburst galaxies, gamma-ray bursts
etc.) Our approach assumes that the number of sources within the GZK
sphere is large enough ($\gtrsim 10^2$) so that the large-scale structure
of the Universe traces well their distribution within $\sim
100$~Mpc. The assumption does not work  for the case of only few sources
and very strong intergalactic magnetic fields (see
Sec.~\ref{sec:mf}).

Interaction of UHE hadrons with cosmic background
radiation limits the propagation distance at high energies; hence, a
limited part of the Universe may contain the sources of UHECRs detected at
the Earth. Non-uniform distribution of matter in this part should reflect
itself in the distribution of the arrival directions of cosmic
rays~\cite{EGmogila}. For this study, the following scheme was used.

\begin{itemize}
 \item
We calculated the density function of the sources
$n(l,b,d)$
as the number density of galaxies
for a given direction (Galactic coordinates $l$, $b$) and distance $d$.
\item
By making use of a numerical propagation
code~\cite{Oleg}, we estimated the fraction $f(d,E_{\rm min})$ of
surviving hadrons with $E>E_{\rm min}$ at the distance $d$ from the source,
assuming either proton or iron injected primaries.
\item
The density of the sources $n(l,b,d)$ was convolved with the survival
function $f(d)$ to obtain the expected distribution of the UHECR arrival
directions.
\end{itemize}

\subsection{The density function of the sources}
\label{sec:density-function}
Most of the previous studies~\cite{EGmogila} used the PSCz
catalog~\cite{PSCz} to reconstruct the source density. However, due to
a limited angular resolution of the IRAS instrument used to compile the
catalog it might not resolve all galaxies in
the regions of high density (clusters)~\cite{2MRS} and thus may
systematically diminish the expected anisotropy due to underestimation of
high number densities of galaxies.
For now, the most complete full-sky catalog of galaxies is the
2MASS XSC~\cite{XSC}. Redshifts are known only for a small fraction of the
2MASS galaxies; the photometric redshifts are proven to be a good measure
of the distance to the rest of them~\cite{photometric-redshifts}. However,
the precision of photometric redshifts is insufficient at low distances;
moreover, artificial nearby sources may appear due to uncertainties in
the determination of the Galactic extinction. We use another full-sky
galaxy catalog, LEDA~\cite{LEDA}, to identify the density function at $d <
30$~Mpc. A volume-limited sample of LEDA~\cite{LEDAcompleteness} for $d <
50$~Mpc is used, and the slice 30~Mpc$< d <50$~Mpc is used for calibration
of the density of the sources (in arbitrary units) determined from LEDA
and from 2MASS XSC photometric redshifts. The latters are calculated
following Ref.~\cite{photometric-redshifts1}, using the Galactic
extinction of Ref.~\cite{GalacticExtinction}, and are used for $d >
30$~Mpc. The details of calculation are given
in~\ref{app:density-function}. Our sample is complete up to the distances
of order 270~Mpc (we assume $H_0=73$~km/s/Mpc).

The actual density of the sources is much less than that of all galaxies.
This means that the density function should be smoothed on a reasonable
scale to suppress unphysical fluctuations. For this study, we smooth
the function $n(l,b,d)$ at the scale of 10~Mpc in $d$ and $\sim 5^\circ$ on
the celestial sphere. The latter value is also motivated by the
uncertainty due to deflection of charged hadrons in the cosmic magnetic
fields (see Sec.~\ref{sec:mf} for discussion) and by the experimental
angular resolution.

\subsection{Calculation of the fraction of surviving hadrons}
\label{sec:survived-fraction}
We use a detailed propagation
code~\cite{Oleg} which is based on kinematic equations written in the
expanding Universe and accounts for numerous interaction processes,
tracing the propagation of nuclei (iron and lighter), nucleons,
gamma-rays, electrons and neutrinos. Parameters of the simulation have
been chosen~\cite{minGZKgamma} in such a way as to provide the best fit to
the HiRes data~\cite{HiRes-cutoff}: for injected protons, the assumed
spectral index at the source is $\alpha =2.55$ and the maximal injected
energy is $E_{\rm max}=1.3 \cdot 10^{21}$~eV; for iron these parameters
are $\alpha = 2.2$ and $E_{\rm max}=1.7 \cdot 10^{22}$~eV,
correspondingly. In both cases, the best fit corresponds to the absence of
the source evolution and to the zero distance to the nearest source. The
same parameters provide reasonable fits to the data of the Pierre Auger
surface detector~\cite{PAO-SDspectrum} if the energy calibration
independent of the fluorescent yield~\cite{Engel-CIC} is used; see
Ref.~\cite{OlegAuger} for other fits to the Auger spectrum.

\begin{figure}
\begin{center}
\includegraphics [width=0.8\textwidth]{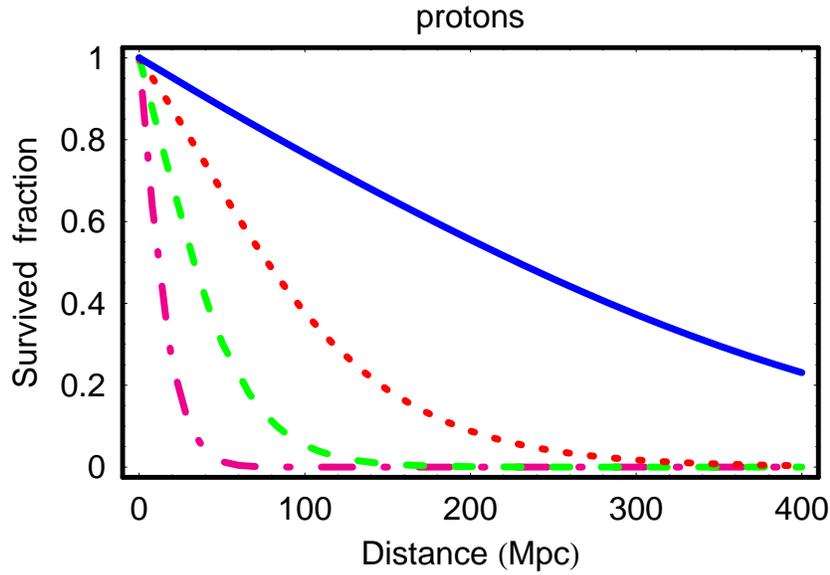}
\end{center}
\caption{
The fraction of surviving hadrons with $E>E_{\rm min}$ as a function of the
distance $d$ from the proton-emitting source
with spectral index $\alpha =2.55$ and a cutoff at $1.3 \cdot 10^{21}$~eV.
Solid line, $E_{\rm min}=4 \cdot 10^{19}$~eV; dotted line, $E_{\rm min}=7\cdot
10^{19}$~eV; dashed line, $E_{\rm min}=10^{20}$~eV; dash-dotted line,
$E_{\rm min}=2 \cdot 10^{20}$~eV.}
\label{fig1p}
\end{figure}

\begin{figure}
\begin{center}
\includegraphics [width=0.8\textwidth]{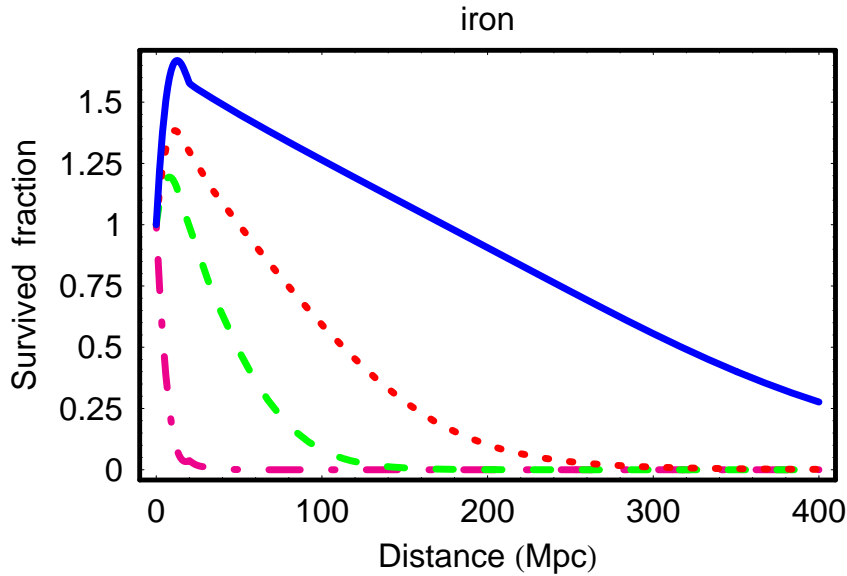}
\end{center}
\caption{
The fraction of surviving hadrons with $E>E_{\rm min}$ as a function of the
distance $d$ from the iron-emitting source
with spectral index $\alpha =2.2$ and a cutoff at $1.7 \cdot 10^{22}$~eV.
Solid line, $E_{\rm min}=4 \cdot 10^{19}$~eV; dotted line, $E_{\rm min}=7\cdot
10^{19}$~eV; dashed line, $E_{\rm min}=10^{20}$~eV; dash-dotted line,
$E_{\rm min}=2 \cdot 10^{20}$~eV.
Note that the {\it number of particles} is first enhanced at dosens
Megaparsecs from the source because of prompt disintegration of heavy
nuclei into a number of light ones and only then becomes suppressed due to
the GZK effect.}
\label{fig1f}
\end{figure}

The functions $f(d,E_{\rm min})$ are
plotted in Figs.~\ref{fig1p},~\ref{fig1f} for different injected primaries
and $E_{\rm min}$.
Since our sample is complete up to $d\sim 270$~Mpc, we see
(cf.~Fig.~\ref{fig1p}, \ref{fig1f}) that the study makes sence for $E
\gtrsim 7\cdot 10^{19}$~eV.

\begin{figure}
\begin{center}
\includegraphics [width=0.8\textwidth]{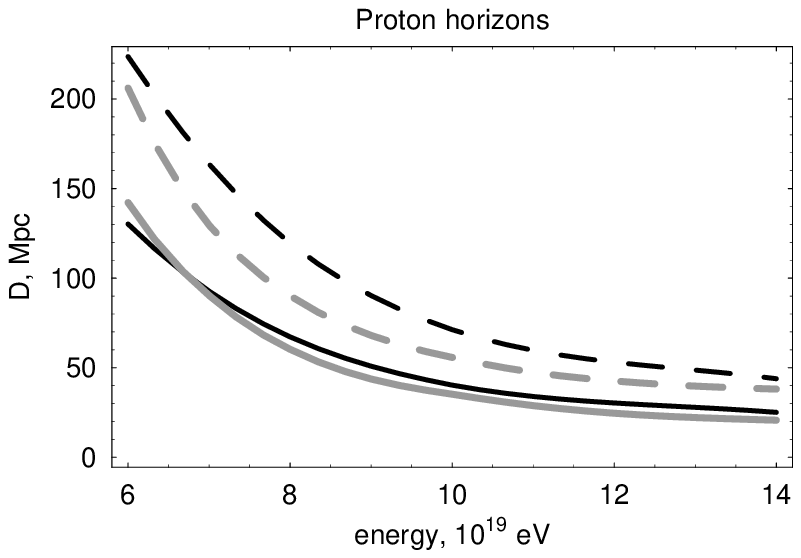}
\end{center}
\caption{
Horizons of protons: proton-injecting sources at distances larger than $D$
contribute a fraction $k$ of the observed flux. Solid lines, $k=0.3$;
dashed lines, $k=0.1$. Black lines, this study; gray lines,
Ref.~\cite{horizon}. Injection spectral index $\alpha =2.7$.
\label{fig:horizonP}
}
\end{figure}

\begin{figure}
\begin{center}
\includegraphics [width=0.8\textwidth]{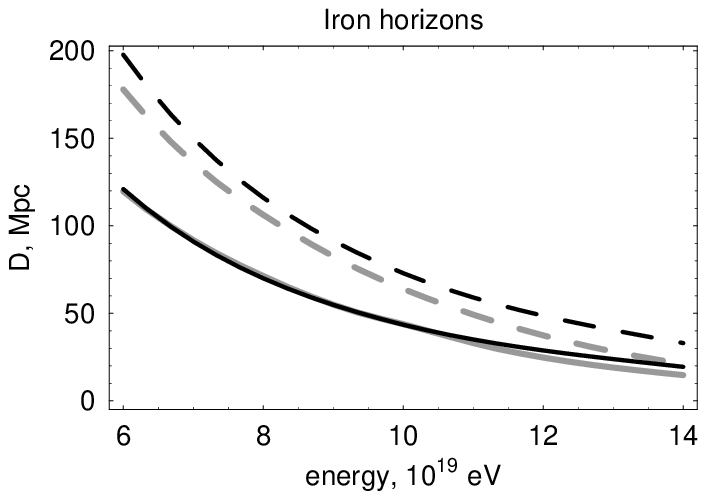}
\end{center}
\caption{
Same as Fig.~\ref{fig:horizonP} but for iron nuclei injected with the
spectral index $\alpha =2.2$.
\label{fig:horizonFe}
}
\end{figure}

These results are compared to a related study~\cite{horizon} in
Figs.~\ref{fig:horizonP}, \ref{fig:horizonFe}.
The agreement between the two simulations is reasonably good though ours
predict slightly larger contribution from distant sources.

\subsection{Effect of the cosmic magnetic fields}
\label{sec:mf}
The previous discussion did not take into account the effect of the cosmic
magnetic fields on the propagation of charged cosmic particles. This
effect is determined by bending of particle trajectories and is therefore
twofold: firstly, the trajectories become longer and secondly, they do not
point back to the origin. The first effect may have important consequences
for the spectrum if the change in the trajectory length becomes comparable
to the attenuation length of the particle; the second effect may change
the expected skymap significantly.

For our purposes, it would be useful to separate the effects of the
extragalactic magnetic fields from those of the magnetic field of the
Milky Way.

\subsubsection{Extragalactic magnetic fields.}
\label{sec:IGMF}

Direct information about the strength of intergalactic magnetic fields is
not available and the corresponding estimates vary by orders of magnitude.
Magnetic fields of order $10^{-11}$~G in the most part of the local
Universe (except dense clusters of galaxies) are predicted by simulations of
Ref.~\cite{DolagMF} while fields as strong as $10^{-8}$~G in voids are
advocated in Ref.~\cite{SiglMF}. Detailed calculations of cosmic-ray
trajectories were given e.g.\ in Refs.~\cite{DolagMF,SiglMF} for both
cases. A typical deflection angle $\chi$ of a particle with energy $E$ and
charge $Z$ may be roughly estimated as
$$
\chi \lesssim 0.1^\circ \frac{7 \cdot 10^{19}~{\rm eV}}{E}
\frac{B}{10^{-11}~G}Z,
$$
where $B$ is the typical magnetic-field strength in the voids. We see that
for the strong-field scenario ($B\sim 10^{-8}$~G), cosmic-ray trajectories
are bend strongly and become entangled ($\chi \sim 360^\circ$) even for
protons ($Z=1$) of $E\sim 7\cdot 10^{19}$~eV (in this case, even a few
sources within $\sim 100$~Mpc would not result in significant anisotropy
of arrival directions).  For less extreme values of $B \lesssim
10^{-9}$~G, typical deflections of protons do not exceed a few degrees at
the super-GZK energies so that the analysis of this work remains
applicable. For the weak-field scenario of Ref.~\cite{DolagMF} which we
assume in what follows, deflections of nuclei as heavy as iron are of
order of, or less than the experimental angular resolution at $E \gtrsim
7\cdot 10^{19}$~eV.

The effect of the lengthening of trajectories on the spectrum of various
nuclei was studied in detail in Ref.~\cite{nucl-spec}; it may be neglected
within our precision for $B<10^{-9}$~G and the energies of interest (see
e.g.\ Fig.~10 of Ref.~\cite{nucl-spec}).

\subsubsection{Galactic magnetic fields.}
\label{sec:GMF}

Much stronger ($\sim 10^{-6}$~G) and somewhat better studied magnetic
fields are present in the Milky Way (see e.g.\ Refs.~\cite{MFmogila} for
reviews of various models with applications to UHECR deflections).
However, their strength is partially compensated by the relatively small
size they occupy (strong fields are localized within the Galactic disk and
possibly in the central regions of the bulge). At $E \gtrsim 7\cdot
10^{19}$~eV, the deflections are always small ($\sim 1^\circ$) for protons
but may be important for heavier nuclei, at least in the directions along
the Galactic plane. The actual skymap at low Galactic latitudes $b$
depends strongly on the particular model of the magnetic-field structure;
however at relatively high $|b| \gtrsim 50^\circ$ the deflections are
always small and the pattern of cosmic structures at these latitudes
(where e.g.\ the Virgo supercluster is located) remains unchanged. The
lengthening of particle trajectories (except possibly for the rare ones
which hit the central region of the Galaxy and may get entangled) is
negligible as compared to the attenuation length.

To summarize this section, the results of our study are not significantly
affected by the account of intergalactic magnetic fields of $B \lesssim
10^{-9}$~G for protons and $B \lesssim 10^{-10}$~G for heavier nuclei and
by the account of the Galactic magnetic fields -- everywhere for protons
and at high Galactic latitudes, $|b| \gtrsim 50^\circ$, for heavy nuclei.
The actual account of these magnetic fields is strongly model-dependent
and is beyond the scope of this paper.

\section{The ``top-down'' scenario: sources follow the distribution of
the Galactic dark matter}
\label{sec:DM}
In ``top-down'' models, UHECRs originate from decays of superheavy
particles (the latters themselves may be produced in decays or collisions
of the topological defects). The distribution of sources thus follows the
distribution of the initial particles; for many realistic models it
corresponds to the distribution of the dark matter. Decays of particles
from the Milky-Way halo dominate the cosmic-ray flux at high energies in
this case. At lower energies, this flux should be supplemented by a
contribution of astrophysical sources to explain the observed spectrum.
Due to a non-central position of the Sun in the Galaxy, the flux should be
higher from the direction of the Galactic center than from the opposite
one.

Decays of the SHDM particles may be described in a more or less
model-independent way because the most important physical phenomenon of
relevance is hadronization which involves light particles and is well
understood. Denote $x \equiv \frac{2E}{M_X}$, where $E$ is the energy of a
decay product of the SHDM particle with mass $M_X$. Then for $10^{-4}
\lesssim x \lesssim 0.1$, spectra of the decay products calculated by
various methods~\cite{ABK1,DMspectra} are in a good agreement with each
other; moreover, the shape of the spectral curve $\frac{dN}{dE}(x)$ does
not depend on $M_X$~\cite{ABK1}. For this study, we use the spectra of
decay products from Ref.~\cite{ABK1}.\footnote{We thank M.~Kachelrie\ss\
for providing numerical tables of the functions caculated there.}

The SHDM decay rate is determined by the concentration $n_X$ and lifetime
$\tau _X$ of the SHDM particles, $\dot{n_X}=n_X/\tau _X$. The flux of
secondary particles at the Earth is then determined by
$$
j={\cal N} \frac{1}{\tau _X}\,\frac{dN}{dE},
$$
where
\begin{equation}
{\cal N}=\int \!d^3r \, \frac{n_X({\bf r})}{4 \pi  r^2}
\label{*}
\end{equation}
is the geometrical factor (${\bf r}$ is the radius-vector from the Earth;
though in principle one should integrate over the Universe and account for
relativity effects, in most interesting cases the dominant contribution
comes from the Galactic halo~\cite{DubTin}). The mass $M_X$ is subject to
cosmological limits (see e.g.\ Ref.~\cite{newKolb} and references
therein); the lifetime $\tau _X$ is much less restricted.

Since the decays of SHDM particles may describe only the highest-energy
part of the spectrum, the flux is assumed to be a sum of two components,
one of which, $F_{\rm EG}$, corresponds to the ``bottom-up'' contribution
(we use the results of Sec.~\ref{sec:EG} to model it) while the second
one, $F_{\rm DM}$, is due to SHDM decays. Neglecting the contribution of
SHDM outside the Milky-Way halo, one may rewrite the integral in
Eq.~(\ref{*}) in the variables related to the Galaxy, namely the distance
$R$ to the Galactic center and the angle $\theta$ between the directions
``observer -- Galactic center'' and ``observer -- integration point''.
This is useful because the distribution of the dark matter in the Galaxy
depends on $R$. Though the actual shape of the distribution is currently a
subject of debates, the profile
$$
n(R)={n_0 \over
\left(\frac{R}{R_S} \right)^\alpha
\left(1+\frac{R}{R_S} \right)^{3-\alpha }
},
$$
is often used,
where $R_S=45$~kpc and $\alpha =1$ for a popular NFW model~\cite{NFW}. We
take the size of the halo $R_h=100$~kpc and the distance from the Sun to
the Galactic center $r_\odot=8.5$~kpc to obtain
$$
\int \! d^3R\, n_X(r) =
\left\{
\begin{array}{ll}
2 I_1 + I_2, & 0\le \theta \le \pi /2,\\
I_2, & \pi /2 \le \theta \le \pi,
\end{array}
\right.
$$
where
$$
I_1=\int\limits_{r_\odot \sin \theta}^{r_\odot} n_1(R)\,dR,
~~~~
I_2=\int\limits_{r_\odot}^{R_h} n_1(R)\,dR,
~~~~
n_1(R)={n(R) \over \sqrt{R^2-r_\odot^2 \sin^2\theta}}
$$
(note that the corresponding equation of Ref.~\cite{ABK2} works only for
$0 \le \theta \le \pi /2$).

Each particular experiment has its own direction-dependent exposure
$A(l,b)$, so the observed flux is a convolution of $F_{\rm EG}+F_{\rm DM}$
with $A$. For instance, AGASA sees the Northern sky and has higher
exposure towards the Galactic anticenter while Southern experiments (SUGAR
and PAO) have higher exposures towards the Galactic center. To simulate
the expected distribution of events, we fit the observed AGASA spectrum
with the sum of $F_{\rm EG}$ and $F_{\rm DM}$ and find the 95\% CL regions
for the three parameters: the coefficients at $F_{\rm EG}$ and $F_{\rm
DM}$ (the latter is related to $\tau _X$) and $M_X$. We impose
additional constraints on the parameters. First, $M_X$
should satisfy the bounds of Ref.~\cite{newKolb} and should be at least
twenty times larger than the energy corresponding to the last point in the
spectrum being fitted (in order to always be in a safe region $x \lesssim
0.1$); these bounds are not very restrictive. However, we impose the
limits on the gamma-ray fraction as reported in
Refs.~\cite{gamma,Yak-gamma,PAO-gamma} which exclude a significant part of
the parameter space. Details of the fitting will be reported elsewhere;
here we will present the results for two particular fits: the best fit for
the AGASA spectrum (which is marginally consistent with gamma limits) and
the fit with the minimal contribution from SHDM (consistent with gamma
limits automatically) still allowed within 95\% CL region of parameters.
We note that the results for the best fit of the PAO
surface-detector spectrum~\cite{PAO-SDspectrum} (calibrated as suggested
in Ref.~\cite{Engel-CIC}) lay between these two.

\section{Predictions for space-based experiments}
\label{sec:experiments}
\subsection{Experiments}
\label{sec:detectors}
The space detector TUS~\cite{TUS} is under construction and is planned to
be launched in 2010. The TUS lower energy threshold is estimated as $7\cdot
10^{19}$~eV (with almost energy-independent sensitivity at higher
energies). The UHECR arrival direction accuracy is different for
different zenith angles $\theta$: for $\theta<30^\circ$, the direction is
estimated only roughly as being vertical in the error cone of 30$^\circ$;
for $30^\circ\lesssim\theta\lesssim60^\circ$ the error approaches
10$^\circ$ and for $60^\circ\lesssim\theta\le90^\circ$ the error is less
than 10$^\circ$. We used the events with
$30^\circ\lesssim\theta\lesssim 90^\circ$ in our estimates; the exposure
factor per one year of operation is about 3000~km$^2$~sr for these cuts.
The TUS direction-dependent exposure $A(l,b)$ is determined by the
satellite orbit parameters and by the zenith angle cut. It can be
conveniently parametrized by a sum of a monopole and a quadrupole over the
celestial sphere.

The JEM-EUSO detector~\cite{EUSO} (to be installed in 2012 on board of the
International Space Station, ISS) is expected to have an instantaneous
aperture of $\sim 6 \cdot 10^5$~km$^2$~sr and a duty cicle of $\sim 20\%$.
JEM-EUSO will be sensitive to cosmic rays with energies $E \gtrsim 4 \cdot
10^{19}$~eV; the energy-dependent sensitivity is reported in
Ref.~\cite{EUSO}. The angular resolution of the detector is a few degrees;
the ISS orbit parameters result in an almost uniform exposure over the sky
(we therefore assume $A(l,b)=$const for JEM-EUSO).

S-EUSO~\cite{S-EUSO} is a proposed detector with an instantaneous aperture
of $\sim 2\cdot 10^6$~km$^2$~sr. The proposal has been submitted to the
ESA ``Cosmic Vision'' program; the proposed launch date is 2017. For
S-EUSO we assume the duty cycle, energy-dependent sensitivity and $A(l,b)$
similar to those simulated for JEM-EUSO.

\subsection{Large-scale observables and optimisation}
\label{sec:observables}
To test one of the hypotheses discussed in Sec.~\ref{sec:EG},
\ref{sec:DM}, we divide the sky into three parts: part 1 corresponds to
the strongest expected excess of events over the uniform distribution,
part 2 corresponds to the strongest deficit and part 3 is the rest of the
sky. The shape and size of the parts, as well as the energy range, are
determined from a priory simulations in order to maximize the signal.
The observable of relevance is the difference between numbers of events in
regions 1 and 2, $n_1-n_2 \equiv \Delta$. Though this strategy is best
suited for TUS with its relatively low aperture and precision, it will be
useful for analysis of early JEM-EUSO and S-EUSO data as well, especially
at the highest energies.

Let $p_i$, $i=1,2,3$, be the probability of an event to arrive from the
$i$th region, $p_1+p_2+p_3=1$; let $N$ be the total number of events in
the sample. Then the probability to have $n_1$ events in the region 1 and
$n_2$ events in the region 2 is given by the multinomial distribution,
$$
P_N(n_1,n_2)={N! \over n_1! n_2!(N-n_1-n_2)!} p_1^{n_1} p_2^{n_2}
p_3^{N-n_1-n_2}.
$$
The probability to have $n_1-n_2=\Delta$ is
$$
\begin{array}{rcl}
P(\Delta)&=&
(1-p_1-p_2)^{N+\Delta} p_2^{-\Delta} {\Gamma(N+1) \over
\Gamma(N+\Delta+1) \Gamma(1-\Delta)} \\
&& \times _2 F_1
\left(-{N+\Delta \over 2}, -{N+\Delta-1 \over 2}, 1-\Delta, {4 p_1 p_2
\over (1-p_1-p_2)^2} \right),
\end{array}
$$
where $\Gamma$ and $_2 F_1$ are the Gamma function and the hypergeometric
function, respectively; $-N \le \Delta \le N$.

For each particular choice of the regions 1 and 2 and of the minimal
energy, we calculate $P_N(\Delta)$ as a function of $N$ for the cases of
the expected signal and of the null hypothesis of isotropic cosmic-ray
flux
(see Ref.~\cite{predictions} for a similar study).
 We expect that an experiment is able to confirm/exclude the
hypothesis at the confidence level $\eta$ if
$$
P_N(\Delta_{\rm signal}-\Delta_{\rm null} \ge 0)>1-\eta.
$$
For a given $\eta$, we determine the required number of events $N$.
Minimization of the time required to collect this number of events
(estimated in each case using the spectrum relevant for the model we test)
results in the choice of the optimal regions and energy.

\subsection{Results for the ``bottom-up'' scenario}
\label{sec:results:EG}
\begin{figure}
\begin{center}
\includegraphics [width=0.8\textwidth]{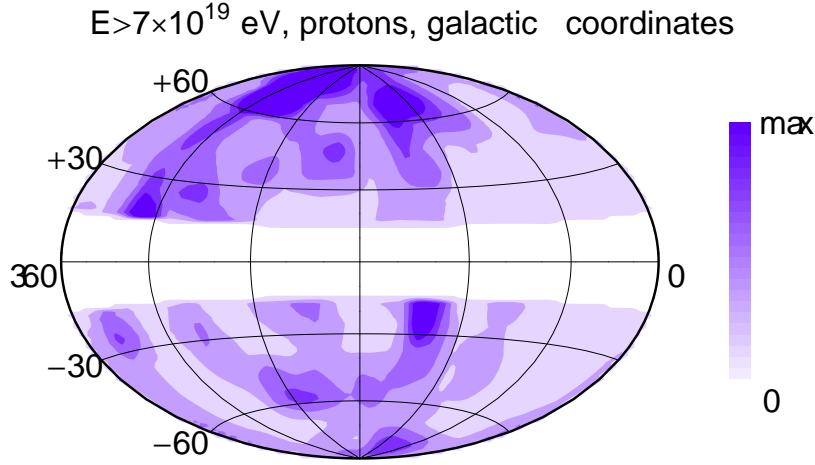}
\end{center}
\caption{
Expected density of events with $E>7 \cdot 10^{19}$~eV from
proton-emitting extragalactic sources plotted on a skymap in galactic
coordinates for uniform exposure. Darker regions correspond to larger
density (linear scale). The white band corresponds to the zone of
avoidance ($|b|<15^\circ$) cut out of the sample.
Numerical values of the expected densities of events for this and other
cases are available online from {\tt http://livni.inr.ac.ru/UHECRskymaps/}.}
\label{fig:skymapP}
\end{figure}

Figure~\ref{fig:skymapP}
presents the expected density of events for
uniform exposure simulated under conditions described in
Sec.~\ref{sec:EG}. The strongest excess of events is expected along the
Supergalactic plane, in particular from the Virgo cluster and the
Perseus--Pisces Supercluster. Compared to the previous studies, ours
suggests a stronger contrast between local voids and filaments, probably
due to the contrast in the source density function (cf.\ the IRAS--2MASS
difference discussed in Sec.~\ref{sec:density-function}).

\begin{figure}
\begin{center}
\includegraphics [width=0.8\textwidth]{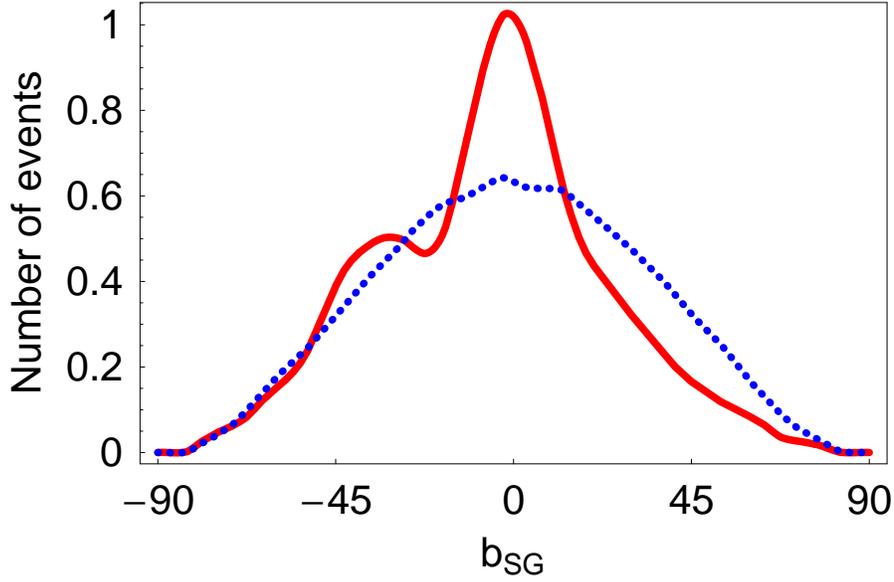}
\end{center}
\caption{
Expected distribution of events (arbitrary units) with $E>7 \cdot
10^{19}$~eV from proton-emitting extragalactic sources (thick line) in the
supergalactic latitude $b_{\rm SG}$, assuming TUS exposure. The dashed line
corresponds to the isotropic distribution.}
\label{fig:l}
\end{figure}

To determine the optimal choice of regions 1 and 2, we tested various cuts
in galactic and supergalactic coordinates and their combinations.
Not surprisingly,
the strongest contrast is achieved in distribution of
events in the supergalactic latitude $b_{\rm SG}$ (illustrated by
Fig.~\ref{fig:l}).
The strongest excess of events corresponds to the band $|b_{\rm
SG}|<15^\circ$ (region 1) while the region of the strongest deficit
(region 2) is the Northern supergalactic hemisphere, $+15^\circ < b_{\rm
SG} <+90^\circ$. For any of the space-based experiments, the optimal
minimal energy of the sample
is $E_{\rm min}=7 \cdot 10^{19}$~eV,
and $\sim 30$ events with these energies are needed in the full-sky sample
for 95\% CL
evidence/exclusion of the hypothesis.

\subsection{Results for the SHDM scenario}
\label{sec:results:DM}
Not surprisingly again, the regions 1 and 2 for the SHDM search are
determined as cones of angle $\psi$ around the Galactic center and
anti-center, respectively. The optimal parameters are given in
Table~\ref{tab:DM} for different experiments.
\begin{table}
\caption{\label{tab:DM}
Optimized search parameters for the SHDM scenario.
The columns
give: (1), experiment name;
(2), assumed SHDM plus extragalactic fit of the AGASA spectrum (see
Sec.~\ref{sec:DM}); (3)--(4), optimized search parameters: $E_{\rm min}$ --
minimal energy of the sample, $\psi$ -- angular radius of the cone around
the Galactic Center (region 1) and Galactic anti-center (region 2);
 (5),
number of events in the sample required for 95\% CL evidence/exclusion of
the hypothesis.
}
\lineup
\begin{indented}
\item[]\begin{tabular}{@{}ccccc}
\br
Experiment&
Fit&
$E_{\rm min}$&
$\psi$&
$N$
\\
(1)&
(2)&
(3)&
(4)&
(5)
\\
\mr
TUS&best&$5\cdot 10^{19}$~eV&70$^\circ$&138\\
&min&$9\cdot 10^{19}$~eV&70$^\circ$&87\\
\mr
JEM-EUSO,&best
&$4\cdot 10^{19}$~eV&70$^\circ$&265\\
S-EUSO&min
&$7\cdot 10^{19}$~eV&60$^\circ$&167\\
\br
\end{tabular}
\end{indented}
\end{table}

\section{Conclusions}
\label{sec:concl}
In this work, we made quantitative predictions for the global anisotropy
of the UHECR arrival directions expected in two distinct scenarios
(``top-down'' and ``bottom-up'') of the origin of the highest-energy
cosmic rays.

Several refinements and improvements resulted in considerable
changes in the predictions as compared to previous studies.
In particular, the patterns in the distribution of arrival directions of
cosmic rays from astrophysical sources are more pronounced than
expected before. The superheavy dark matter scenarios consistent
with current data predict less pronounced anisotropy.

We developed optimal observables for distinction
(exclusion) of these two scenarios. The actual required observational time
depends strongly on currently uncertain duty cycle, fluorescent yield
and the actual spectrum. In any case, some of the scenarios will be
tested even with the limited statistics of the first space-based UHECR
detector, TUS, before (or at the time of) the planned launch of JEM/EUSO.
However, only JEM-EUSO and/or
S-EUSO will be able to detect the SHDM component predicted by the
``minimal'' fit of the AGASA data consistent at 95\% CL with the spectrum
and with photon limits.

To observe, at the 95\% CL,
the patterns correlated with cosmic large-scale structure, one needs $\sim
30$ events in the full-sky sample with energies $E>7 \cdot
10^{19}$~eV. If these patterns show up at high confidence with lower
statistics,
this might indicate either significant
underestimation of the particle energies or a problem in theoretical
understanding of the origin and/or propagation of ultra-high-energy cosmic
particles. One may try to use these patterns as a rough but
independent tool for the energy calibration, quite important for
space-based detectors.

\ack The authors are indebted to D.~Gorbunov, G.~Rubtsov and
D.~Semikoz for discussions. We acknowledge the use of online
tools~\cite{XSC,LEDA,GalacticExtinction}. This work was supported in part
by the grants RFBR 07-02-00820 (OK and ST) and NS-7293.2006.2 (government
contract 02.445.11.7370, OK and ST) and by the Russian Science Support
Foundation fellowship (ST). Simulations of the cosmic-ray propagation have
been performed at the computer cluster of the Theoretical Division of INR
RAS.

\appendix

\section{Construction of the density function of the sources}
\label{app:density-function}
\subsection{The 2MASS sample}
\label{app:2MASSsample}
The most complete full-sky catalog of galaxies, the 2MASS XSC~\cite{XSC},
does not contain distances to objects listed there  (simply because their
redshifts have not been measured in most cases). However, one may use the
``photometric reshifts''~\cite{photometric-redshifts} assuming that the
galaxies are standard candles in the infrared. More precisely, one assumes
that all galaxies have the same absolute magnitudes in $K$-band,
$M_K=M_\star$ (this assumption works good enough in average for large
samples of galaxies, see discussion in Ref.~\cite{photometric-redshifts}).
Following Refs.~\cite{photometric-redshifts,Mstar}, we assume
$M_\star=-24.0$. One has~\cite{photometric-redshifts}
$$
M_K=K_{\rm corr}-5\log\frac{D_L(z)}{r_0}-k(z),
$$
where $K_{\rm corr}$ is the apparent $K$-magnitude corrected for the
Galactic extinction, $D_L(z)$ is the luminosity distance (for redshifts
$z\lesssim 0.1$ which we are interested in, $D_L(z)\approx z/H_0$; we use
the Hubble constant $H_0=73$~km$\cdot$s$^{-1}\cdot$Mpc$^{-1}$),
$r_0=10$~pc and $k(z)$ is the cosmic-reddening correction, $k(z)\approx
-6\log(1+z)$. The distance $d$ to a galaxy is then determined as
$$
d=\frac{5}{12}\left(1-\sqrt{1-\frac{24}{5}H_0 r_0 10^{(K_{\rm
corr}-M_\star)/5}} \right) H_0^{-1}.
$$
We use the model of Ref.~\cite{GalacticExtinction} for correcting the
observed $K$-magnitude $K_{\rm obs}$ for the Galactic extinction:
$$
K_{\rm corr}=K_{\rm obs}-0.367 E(B-V),
$$
where $E(B-V)$ is found for a given direction $(l,b)$ by making use of the
code \verb"dust_getval" available from the
website~\cite{GalacticExtinction}.

According to Ref.~\cite{photometric-redshifts}, the 2MASS XSC is a
complete catalog of galaxies with $|b| \ge 5^\circ$ and $K<13^m$. This
corresponds to the photometric redshift $z\approx 0.066$ (that is,
$d\approx 270$~Mpc).

\subsection{The LEDA sample and matching}
\label{app:LEDA}
The photometric redshifts at low distances are not always reliable because
of their relatively low precision (which affects the results more
significantly for closeby objects). Moreover, the sample of nearby
galaxies determined by photometric redshifts suffer from contamination by
Galactic objects (though firm Galactic identifications are listed in a
catalog available from the XSC website~\cite{XSC} and can easily be
removed, one expects that a fraction of these objects is still present)
and by galaxies for which the galactic extinction was determined
poorly (e.g.\ due to molecular clouds in front of them). Therefore, one
needs another source of data at small distances $d$.

The most complete optical full-sky catalog of galaxies is the LEDA
database~\cite{LEDA}. It contains information about radial velocities
(known mostly for nearby galaxies). The completeness of the catalog has
been studied in Ref.~\cite{LEDAcompleteness} where volume-limited complete
samples were determined. We use the sample of galaxies with known
distances up to 50~Mpc, complete for absolute $B$-magnitudes $M_B<-19 $ and
$|b| \ge 15^\circ$. From this sample, we construct the source density
function $n_{\rm LEDA}(l,b,d)$ and compare it to $n_{\rm XSC}(l,b,d)$ in
the region 30~Mpc$\le d \le 50$~Mpc to determine the relative
normalization of the two functions:
$$
n_{\rm XSC} \simeq 1.03 n_{\rm LEDA}.
$$
The normalization factor is close to unity which reflects the completeness
of the both samples. For our study, we therefore use the function
$$
n(l,b,d)=\left\{
\begin{array}{ll}
1.03 n_{\rm LEDA}(l,b,d),& d\le 30~{\rm Mpc},\\
n_{\rm XSC}(l,b,d),& 30~{\rm Mpc}<d \le 270~{\rm Mpc}.
\end{array}
\right.
$$
Though $n_{\rm XSC}$ is determined for $|b|\ge 5^\circ$, we have to limit
ourselves to $|b| \ge 15^\circ$ because of the zone-of-avoidance cut in
the LEDA sample.

When the
XSCz catalog~\cite{XSCz} (see also Ref.~\cite{2MRS}) containing distances
to many 2MASS galaxies will become available, the use of the separate
database for nearby galaxies will be not necessary and it will be possible
to include the region $5^\circ \le |b| <15^\circ$ in the study.

\begin{figure}
\begin{center}
\includegraphics [width=0.8\textwidth]{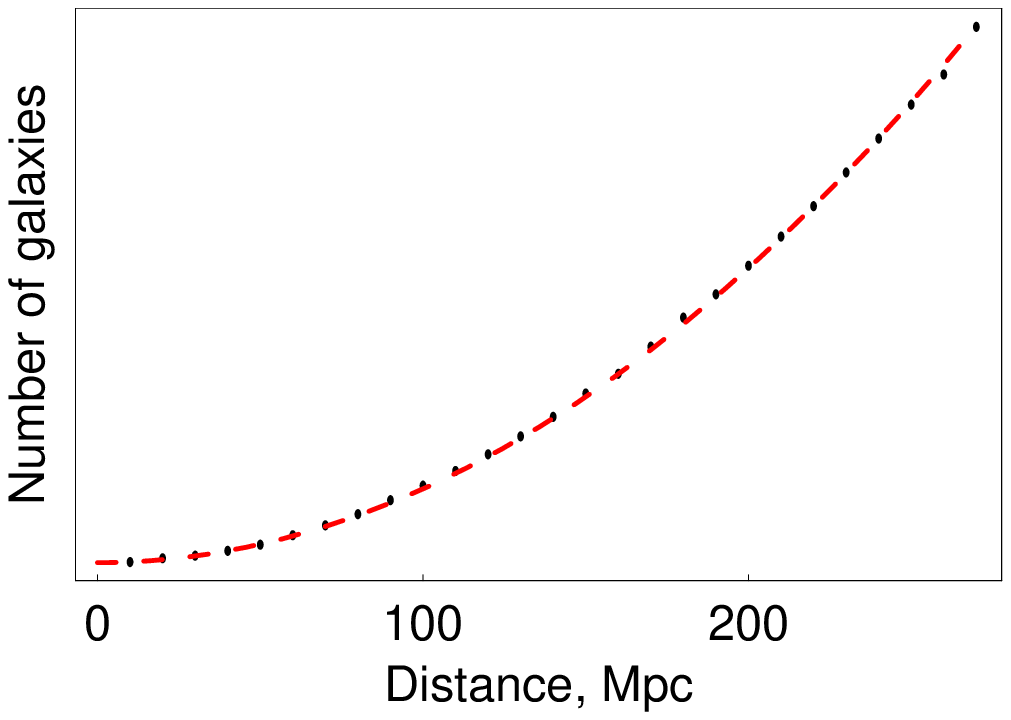}
\end{center}
\caption{
The number of galaxies in our sample in slices of 10~Mpc in distance, as
a function of distance. The dashed line represents the quadratic fit
expected for a complete sample.}
\label{fig:completeness}
\end{figure}

The total number of galaxies in our sample is 212700.
The
completeness of our catalog is illustrated by Fig.~\ref{fig:completeness}.


\section*{References}

\begin{thebibliography}{47}

\bibitem{EGmogila}
Waxman E, Fisher K B and Piran T,
{\it The signature of a correlation between $>10^{19}$--eV cosmic ray
sources and large scale structure,} 1997 {\it  Astrophys.\ J.}\ {\bf 483}
  1 [arXiv:astro-ph/9604005];
Evans N W, Ferrer F and Sarkar S,
  {\it The anisotropy of the ultra-high energy cosmic rays,} 2002
{\it    Astropart.\ Phys.}\
  {\bf 17}  319
  [arXiv:astro-ph/0103085];
Smialkowski A, Giller M and Michalak W,
  {\it Luminous infrared galaxies as possible sources of the UHE cosmic
  rays,}                   2002
{\it    J.\ Phys.\ G}
  {\bf 28} 1359
  [arXiv:astro-ph/0203337];
Sigl G, Miniati F and Ensslin T A,
{\it Ultra-high energy cosmic ray probes of large scale structure and
  magnetic
fields,}                   2004
{\it    Phys.\ Rev.\  D}
  {\bf 70} 043007
  [arXiv:astro-ph/0401084];
Cuoco A {\it et al.},
 {\it The footprint of large scale cosmic structure on the ultra-high
  energy
cosmic ray distribution,} 2006
{\it    JCAP}
  {\bf 0601} 009
  [arXiv:astro-ph/0510765]
\bibitem{GZK}
Greisen K,
{\it End To The Cosmic Ray Spectrum?,} 1966
{\it    Phys.\ Rev.\ Lett.}\
  {\bf 16} 748;
Zatsepin G T and Kuzmin V A,
{\it Upper limit of the spectrum of cosmic rays,} 1966
{\it    JETP Lett.}\
  {\bf 4} 78
  [{\it  Pisma Zh.\ Eksp.\ Teor.\ Fiz.}\  {\bf 4} 114]
\bibitem{TD}
Berezinsky V, Kachelrie\ss\ M and Vilenkin A,
{\it Ultra-high energy cosmic rays without GZK cutoff,} 1997
  {\it  Phys.\ Rev.\ Lett.}\
  {\bf 79} 4302
[arXiv:astro-ph/9708217];
Kuzmin V A and Rubakov V A,
{\it Ultrahigh-energy cosmic rays: A window on postinflationary
  reheating epoch
of the universe?,}       1998
{\it    Phys.\ Atom.\ Nucl.}\
  {\bf 61} 1028
  [{\it  Yad.\ Fiz.}\
  {\bf 61} 1122]
  [arXiv:astro-ph/9709187]
\bibitem{TDrev}
Bhattacharjee P and Sigl G,
{\it Origin and propagation of extremely high energy cosmic rays,} 2000
{\it    Phys.\ Rept.}\
  {\bf 327} 109
[arXiv:astro-ph/9811011]
Kachelrie\ss\ M,
{\it Status of particle physics solutions to the UHECR puzzle,} 2004
{\it   Comptes Rendus Physique}
  {\bf 5} 441
\bibitem{DubTin}
Dubovsky S L and Tinyakov P G,
{\it Galactic anisotropy as signature of CDM-related ultra-high energy  cosmic
rays,}   1998
{\it    JETP Lett.}\
  {\bf 68} 107
  [arXiv:hep-ph/9802382]
\bibitem{DMmogila}
Berezinsky V and Mikhailov A,
  {\it Anisotropy of ultra high energy cosmic rays in the dark matter halo
 models,}               1999
{\it    Phys.\ Lett.\  B}
  {\bf 449}, 237
  [arXiv:astro-ph/9810277];
Medina-Tanco G A and Watson A A,
{\it Dark matter halos and the anisotropy of ultra-high energy cosmic
rays,} 1999
  {\it    Astropart.\ Phys.}\ {\bf 12} 25
  [arXiv:astro-ph/9903182]
\bibitem{ABK1}
Aloisio R, Berezinsky V and Kachelrie\ss\ M,
{\it Ultra high energy cosmic rays spectra in top-down models,} 2004
{\it    Nucl.\ Phys.\ Proc.\ Suppl.}\
  {\bf 136} 319
  [arXiv:astro-ph/0409222]
\bibitem{ABK2}
Aloisio R, Berezinsky V and Kachelrie\ss\ M,
{\it On the status of superheavy dark matter,} 2006
{\it    Phys.\ Rev.\  D}
  {\bf 74} 023516
  [arXiv:astro-ph/0604311]
\bibitem{AGASAspectrum}
Takeda M {\it et al.},
{\it Energy determination in the Akeno Giant Air Shower Array experiment,}
2003 {\it    Astropart.\ Phys.}\
  {\bf 19} 447
  [arXiv:astro-ph/0209422]
\bibitem{HiRes-cutoff}
Abbasi R {\it et al.},
(HiRes Collaboration),
{\it Observation of the GZK cutoff by the HiRes experiment,}
  arXiv:astro-ph/0703099.
\bibitem{PAO-SDspectrum} Roth M
{\it et al.} (Pierre Auger Collaboration),
{\it Measurement of the UHECR energy spectrum using data from the Surface
Detector of the Pierre Auger Observatory,}
Proc. 30th ICRC (2007),
  arXiv:0706.2096 [astro-ph].
\bibitem{Engel-CIC} Engel R
{\it et al.} (Pierre Auger Collaboration),
{\it Test of hadronic interaction models with data from the Pierre Auger
Observatory,}
Proc. 30th ICRC (2007).
  arXiv:0706.1921 [astro-ph].
\bibitem{gamma}
Rubtsov G I {\it et al.},
{\it Upper limit on the ultra-high-energy photon flux from AGASA and Yakutsk
data,}                                     2006
{\it    Phys.\ Rev.\  D }
  {\bf 73} 063009
\bibitem{Yak-gamma}
Glushkov A V {\it et al.},
{\it Constraining the fraction of primary gamma rays at ultra-high energies from
the muon data of the Yakutsk extensive-air-shower array,} 2007
{\it    JETP Lett.}\
  {\bf 85} 131
[arXiv:astro-ph/0701245]
\bibitem{PAO-gamma} Healy M D
{\it et al.} (Pierre Auger Collaboration),
  {\it Search for Ultra-High Energy Photons with the Pierre Auger
  Observatory,}  Proc. 30th ICRC (2007),
arXiv:0710.0025 [astro-ph].
\bibitem{Risse}
Risse M and Homola P,
{\it Search for ultra-high energy photons using air showers,} 2007
{\it    Mod.\ Phys.\ Lett.\  A}
  {\bf 22} 749
[arXiv:astro-ph/0702632]
\bibitem{PAOagn}
Abraham J {\it et al.}  [Pierre Auger Collaboration], 2007,
{\it Correlation of the highest energy cosmic rays with nearby
extragalactic objects}, Science {\bf 318} 938
  [arXiv:0711.2256].
\bibitem{Comment}
Gorbunov D {\it et al.}, 2007,
{\it Comment on 'Correlation of the Highest-Energy Cosmic Rays
with Nearby Extragalactic Objects'}, arXiv:0711.4060.
\bibitem{TUS}
Abrashkin V I {\it et al.},
  {\it
The TUS space fluorescence detector for study of UHECR and other phenomena
  of variable fluorescence light in the atmosphere}, 2006
{\it    Adv.\ Space Research}
  {\bf 37} 1876;
Abrashkin V {\it et al.},
{\it  Status of the TUS space detector preparation for UHECR study},
  Proc. 30th ICRC (2007).
\bibitem{EUSO}
Ebisuzaki T {\it et al.},
{\it The JEM-EUSO Mission},
  Proc. 30th ICRC (2007).
\bibitem{S-EUSO}
Santangelo A {\it et al.},
{\it S-EUSO: a
proposal for a space-based observatory of Ultra-high-Energy cosmic
particles}, submitted to the ESA Cosmic Vision 2015--2025 Program (2007).
\bibitem{EGRET-statistics}
Mattox J R  {\it et al.},
{\it The Likelihood Analysis of EGRET Data}, 1996
{\it  Astrophys.\ J.}\
  {\bf 461} 396;
Tompkins W,
  {\it Applications of likelihood analysis in gamma-ray astrophysics,}
  arXiv:astro-ph/0202141;
Gorbunov D S  {\it et al.},
  {\it Identification of extragalactic sources of the highest energy EGRET
  photons by correlation analysis,} 2005
{\it    Mon.\ Not.\ Roy.\ Astron.\ Soc.\ Lett.}\
  {\bf 362} L30
  [arXiv:astro-ph/0505597]
\bibitem{Sommers}
Sommers P,
{\it Cosmic Ray Anisotropy Analysis with a Full-Sky Observatory,} 2001
{\it    Astropart.\ Phys.}\
  {\bf 14} 271
  [arXiv:astro-ph/0004016]
\bibitem{sources}
Torres D F and Anchordoqui L A,
{\it Astrophysical origins of ultrahigh energy cosmic rays,} 2004
{\it    Rept.\ Prog.\ Phys.}\
  {\bf 67} 1663
  [arXiv:astro-ph/0402371];
Gorbunov D and Troitsky S,
{\it A comparative study of correlations between arrival directions of
ultra-high-energy cosmic rays and positions of their potential  astrophysical
sources,} 2005
{\it    Astrop.\ Phys.}\
  {\bf 23} 175
 [arXiv:astro-ph/0410741]
\bibitem{Oleg}
Kalashev O E, Kuzmin V A and Semikoz D V,
{\it Ultra high energy cosmic rays propagation in the galaxy and  anisotropy,}
2001 {\it    Mod.\ Phys.\ Lett.\  A}
  {\bf 16} 2505
[arXiv:astro-ph/0006349];
Kalashev O E, Kuzmin V A and Semikoz D V,
{\it Top-down models and extremely high energy cosmic rays,}
  arXiv:astro-ph/9911035.
\bibitem{PSCz}
Saunders W {\it et al.},
{\it The PSCz Catalogue,} 2000
{\it    Mon.\ Not.\ Roy.\ Astron.\ Soc.}
  {\bf 317} 55
  [arXiv:astro-ph/0001117]
\bibitem{2MRS}
Huchra J {\it et al.},
{\it The 2MASS redshift survey,}
http://cfa-www.harvard.edu/$\tilde{~}$huchra/2mass/
\bibitem{XSC}
Jarrett T H  {\it et al.},
{\it 2MASS Extended Source Catalog: Overview and Algorithms,} 2000
{\it    Astron.\ J.}\
  {\bf 119} 2498
 [arXiv:astro-ph/0004318];
http://irsa.ipac.caltech.edu/cgi-bin/Gator/nph-dd?catalog=fp$\underline{~}$xsc
\bibitem{photometric-redshifts}
Jarrett T,
{\it Large Scale Structure in the Local Universe: The 2MASS Galaxy Catalog,}
  arXiv:astro-ph/0405069.
\bibitem{LEDA}
Paturel G  {\it et al.},
{\it HYPERLEDA. I. Identification and designation of
galaxies}, 2003
{\it  Astron.\ Astrophys.}
  {\bf 412} 45;
http://leda.univ-lyon1.fr
\bibitem{LEDAcompleteness}
Courtois H {\it et al.},
{\it The LEDA galaxy distribution: I. Maps of the Local Universe,} 2004
{\it    Astron.\ Astrophys.}\
  {\bf 423} 27
 [arXiv:astro-ph/0403545]
\bibitem{photometric-redshifts1}
Jarrett T H {\it et al.},
{\it 2MASS Galaxy Colors: Hercules Cluster}, 1998
{\it  Bull.\ Amer.\ Astron.\ Soc.}
  {\bf 30} 901;
Kochanek C S {\it et al.},
{\it Clusters of galaxies in the local universe,} 2003
{\it    Astrophys.\ J.}\
  {\bf 585}, 161
  [arXiv:astro-ph/0208168]
\bibitem{GalacticExtinction}
Schlegel D J, Finkbeiner D P and Davis M,
{\it Maps of Dust IR Emission for Use in Estimation of Reddening and CMBR
Foregrounds,}          1998
{\it    Astrophys.\ J.}\
  {\bf 500} 525
  [arXiv:astro-ph/9710327];
http://astro.berkeley.edu/dust/index.html
\bibitem{minGZKgamma}
Gelmini G, Kalashev O and Semikoz D V,
{\it GZK photons in the minimal ultra high energy cosmic rays model,}
  arXiv:astro-ph/0702464.
\bibitem{OlegAuger}
Arisaka K
{\it et al.},
{\it Composition of UHECR and the Pierre Auger Observatory Spectrum,}
  arXiv:0709.3390 [astro-ph].
\bibitem{horizon}
 Harari D, Mollerach S and Roulet E,
{\it On the ultra-high energy cosmic ray horizon,} 2006
{\it    JCAP}
  {\bf 0611} 012
[arXiv:astro-ph/0609294]
\bibitem{DolagMF}
Dolag K
{\it et al.},
 {\it
Mapping deflections of Ultra-High Energy Cosmic Rays in Constrained
Simulations of Extragalactic Magnetic Fields,} 2004
  JETP Lett.\  {\bf 79}  583
  [arXiv:astro-ph/0310902];
Dolag K
{\it et al.},
{\it Constrained simulations of the magnetic field in the local universe
and  the propagation of UHECRs,} 2005
  JCAP {\bf 0501} 009
  [arXiv:astro-ph/0410419].
\bibitem{SiglMF}
Sigl G, Miniati F and Ensslin T A,
  {\it Signatures of magnetized large scale structure in ultra-high energy
cosmic rays,}
  arXiv:astro-ph/0309695;
Sigl G, Miniati F and Ensslin T A,
  {\it Cosmic magnetic fields and their influence on ultra-high energy
cosmic ray propagation,} 2004
  Nucl.\ Phys.\ Proc.\ Suppl.\  {\bf 136} 224
  [arXiv:astro-ph/0409098].
\bibitem{nucl-spec}
Globus N, Allard D and Parizot E,
  {\it Propagation of high-energy cosmic rays in extragalactic turbulent
magnetic fields: resulting energy spectrum and composition,}
  arXiv:0709.1541 [astro-ph].
\bibitem{MFmogila}
Tinyakov P G and Tkachev I I,
{\it Tracing protons through the galactic magnetic field: A clue for charge
composition of ultrahigh-energy cosmic rays,} 2002
Astropart.\ Phys.\  {\bf 18} 165
[arXiv:astro-ph/0111305];
Prouza M and Smida R,
  {\it The Galactic magnetic field and propagation of ultra-high energy
  cosmic rays,} 2003 Astron.\ Astrophys.\ {\bf 410} 1
  [arXiv:astro-ph/0307165];
Troitsky S V,
  {\it Magnetic deflections and possible sources of
clustered ultra-high-energy cosmic rays,} 2006
  Astropart.\ Phys.\  {\bf 26} 325
  [arXiv:astro-ph/0505262];
 Kachelriess M, Serpico P D and Teshima M,
  {\it The Galactic magnetic field as spectrograph for ultra-high energy
  cosmic rays,} 2006
  Astropart.\ Phys.\  {\bf 26} 378
  [arXiv:astro-ph/0510444].
\bibitem{DMspectra}
Sarkar S and Toldra R,
{\it The high energy cosmic ray spectrum from massive particle decay,} 2002
{\it    Nucl.\ Phys.\  B}
  {\bf 621} 495
  [arXiv:hep-ph/0108098];
Barbot C and Drees M,
{\it Production of ultra-energetic cosmic rays through the decay of  super-heavy
$X$ particles,} 2002
{\it    Phys.\ Lett.\  B}
  {\bf 533} 107
  [arXiv:hep-ph/0202072];
Barbot C and Drees M,
{\it Detailed analysis of the decay spectrum of a super-heavy X particle,}
  2003
{\it    Astropart.\ Phys.}\
  {\bf 20} 5
  [arXiv:hep-ph/0211406]
\bibitem{newKolb}
Chung D J H  {\it et al.},
{\it Isocurvature constraints on gravitationally produced superheavy dark
matter,}                2005
{\it    Phys.\ Rev.\  D}
  {\bf 72} 023511
[arXiv:astro-ph/0411468]
\bibitem{NFW}
 Navarro J F, Frenk C S and White S D M,
{\it The Structure of Cold Dark Matter Halos,} 1996
{\it    Astrophys.\ J.}\
  {\bf 462} 563
  [arXiv:astro-ph/9508025]
\bibitem{predictions}
Gorbunov D S  {\it et al.},
{\it Estimate of the correlation signal between cosmic rays and BL Lacs in
future data,}             2006
{\it    JCAP}
  {\bf 0601} 025
  [arXiv:astro-ph/0508329]
\bibitem{Mstar}
Cole S {\it et al.}  (The 2dFGRS Collaboration),
{\it The 2dF Galaxy Redshift Survey: Near Infrared Galaxy Luminosity
Functions,} 2001
{\it    Mon.\ Not.\ Roy.\ Astron.\ Soc.}\
  {\bf 326} 255
[arXiv:astro-ph/0012429];
Kochanek C S {\it et al.},
{\it The K-Band Galaxy Luminosity Function,} 2001
{\it    Astrophys.\ J.}\
  {\bf 560} 566
[arXiv:astro-ph/0011456]
\bibitem{XSCz}
Jarrett T,
{\it 2MASS Galaxy Redshift Catalog (XSCz),}\\
http://spider.ipac.caltech.edu/staff/jarrett/XSCz/index.html

\end{thebibliography}
\end{document}